\documentclass[preprint]{JHEP3}

\usepackage{epsfig}     
\usepackage{cancel}

\newcommand{\beq}{\begin{equation}}
\newcommand{\eeq}{\end{equation}}
\newcommand{\beqa}{\begin{eqnarray}}
\newcommand{\eeqa}{\end{eqnarray}}
{

\title{Topics in Nonlinear Sigma Models in $D=3$}

\author{Ko Furuta \\
Department of Physics, National Tsing Hua University, 
101 Section 2 Kuang Fu Road, Hsinchu, 
Taiwan 300, R.O.C.\\ 
Email: \email{furuta@phys.nthu.edu.tw}}
\author{\speaker{Takeo Inami}
        \\
        Department of Physics, Chuo University, 
, Kasuga, Bunkyo-ku, Tokyo 112-8551, Japan\\
        E-mail: \email{inami@phys.chuo-u.ac.jp}}
\author{Masayoshi Yamamoto \\
Department of Physics, Ewha Womans University,
Seoul 120-750, Korea\\ 
E-mail: \email{yamamoto@dante.ewha.ac.kr}}

\abstract{Nonlinear sigma models (NLSM) in $d=3$ have many 
interesting and non-trivial features, which were explored poorly 
in contrast with NLSM in $d=2$ and $d=4$. We present a few 
results from our study  of the perturbative and non-perturbative 
properties of three-dimensional (3D) NLSM.

i) We have shown that cancellation of ultra-violet (UV) divergences 
takes place in 3D extended ($N= 2, 4$) supersymmetric NLSM 
in low orders of the $1/n$ expansion. 

ii) We consider noncommutative extension of the 3D $CP^n$ 
model, and study low-energy dynamics of BPS solitons in this 
model. We also discuss briefly dynamics of non-BPS solutions.}

\begin{document}
\section{Introduction}
Nonlinear sigma models (NLSM) in various dimensions $d$ have 
applications in many branches of physics. In mathematical physics, 
they provide a class of integrable field theories in $d=2$.
\footnote{Integrability does not hold at quantum level for some of 
the integrable field theories obtained this \\
way , e. g., $CP^n (n \geq 2)$. This remark is due to some of the 
particpants of this conference.} 
Superstrings and super-membranes are formulated as field theories 
in the form of NLSM in $d=2$ and $d=3$, respectively. In 
condensed matter physics, quantum two-dimensional Heisenberg 
model can be  described (approximately) by three-dimensional 
(3D) $O(3)$ NLSM \cite{Halp}.

Low-dimensional (supersymmetric) NLSM share many of the 
important properties with four-dimensional (4D) (supersymmetric) 
Yang-Mills theories, in both perturbative and non-perturbative  
aspects, e.g., the  property of $\beta$-function and the existence 
of instantons/solitons. They provide toy field theories of 4D 
Yang-Mills theories.

One recent development in the theories of string and branes is 
Yang-Mills theories on  noncommutative space (or space-time), 
which arise as low-energy description of string theories 
\cite{CDS,CK,SW}. Noncommutative field theories have also been 
studied, in both their perturbative and non-perturbative aspects, 
for their own right and in a more general context. See, for instance, 
\cite{DougNek, MinwRaamSei}.

In this talk we present a few results from our recent studies of 
NLSM in $d=3$ : \\
i) Cancellation of ultra-violet (UV) divergences in the 3D 
extended-supersymmetric NLSM \cite{ISY1, ISY2} 
(sec.\ref{sec:UV}). \\
ii) The properties of solitons in noncommutative extension of 3D 
NLSM \cite{FINY1, FINY2} (sec.\ref{sec:Scatt} and 
sec.\ref{sec:NonBPS}). \\
iii) In addition, we briefly mention our attempt at noncommutative 
extension of integrable models in $d=2$ (\cite{FurIna} and 
unpublished work) (sec.\ref{subsec:Integrable}).

\bigskip
\section{UV properties of supersymmetric nonlinear sigma 
models in $d=3$} 
\label{sec:UV}

NLSM in different space-time dimensions $d$ have distinct 
ultra-violet (UV) properties. NLSM are renormalizable in $d=2$, 
while they are non-renormalizable for $d \geq4$. In between, 
i.e., at $d=3$, (non)renormalizability of NLSM is a subtle 
question. They are apparently non-renormalizable in perturbation 
(in the coupling constant). It was shown some time ago that 3D 
NLSM (e.g., $O(n), CP^n$ models) are renormalizable if the 
theory is defined in the  $1/n$ expansion \cite{Arefetc, RWP}.

It is known that supersymmetric (SUSY) field theories have weaker 
UV divergences than their bosonic counterparts. Good examples 
are 2D NLSM and 4D super-Yang Mills theories. 2D Bosonic NLSM 
are renormalizable, while those with $N=2$ extended 
SUSY are finite up to four loops (in the case of Ricci flat target 
space) \cite{GvdVZ} and those with $N=4$ extended SUSY are 
all-loop finite. The $\beta$-function of 4D super-Yang-Mills 
theories is one-loop exact in the $N=2$ SUSY case and vanishes in 
the $N=4$ case. The question naturally arises whether there is 
some class of SUSY field theories in $d=3$ in which similar 
cancellation of UV divergences takes place.

In view of this we have studied the UV divergence properties of 
3D NLSM with extended supersymmetry  and found that this class 
of field theories has UV divergence properties similar to those of 
2D NLSM and 4D Yang-Mills theories. UV divergences manifest 
themselves in $\beta$-functions. We present the 
$\beta$-functions of 3D NLSM obtained in 
\cite{ISY1, ISY2}.\\
$N=0, \ N=1 \ O(n)$ \cite{RWP, KourMahaetc}
\beq
\beta(g) = g(1-g/2\pi) + 
next{\textstyle-}to{\textstyle-}leading \ correc.
\eeq
$N=2 \ CP^n$ \cite{ISY1, CiuGra}
\beq
\beta(g) = g(1-g/2\pi) + \cancel{next{\textstyle-}to
{\textstyle-}leading \ correc}.
\eeq
$N=4$\,\cite{ISY2}
\beq
\beta(g) = g + \cancel{leading \ correc} + 
\cancel{next{\textstyle-}to{\textstyle-}leading \ correc}.
\eeq
Here, \cancel{term} denotes that the term vanishes.
The $\beta$-function in the  $N=2$ and $N=4$ cases are 
shown in fig.\ref{beta}. 

\FIGURE[t]{
\parbox{160mm}{
\begin{center}
\epsfig{file=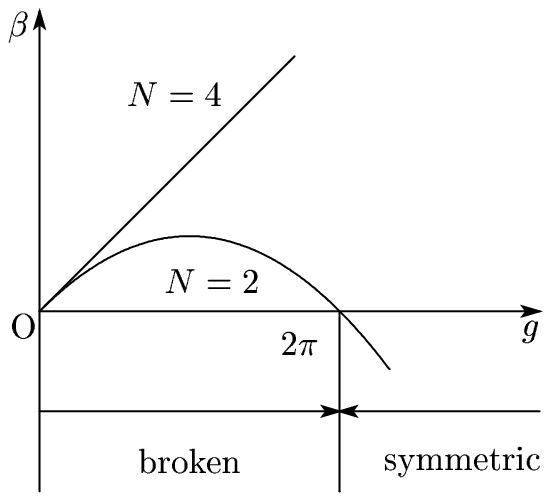}
\end{center}}
\caption
{$\beta$-function of $N=2$ and $N=4$ SUSY NLSM. 
The critical point is at $g=g_{\rm C}=2\pi$ in the $N=2$ 
case, as indicated in the figure. There are two phases, "broken" 
($g < g_{\rm C}$) and "symmetric" ($g > g_{\rm C}$), in the 
$N=0, 1$, and $N=2$ cases.}} 
\label{beta}

Remarkably, next-to-leading order corrections to the 
$\beta$-function vanish in $N=2$ SUSY NLSM, while both 
leading and next-to-leading order corrections vanish in $N=4$ 
SUSY NLSM. These low order results are consistent with the 
possibility that $\beta(g)$ is leading-order exact in the $N=2$ 
case and has no quantum corrections in the $N=4$ case, which 
resembles $\beta(g)$ of 4D super-Yang-Mills theories. It is very 
interesting to see whether this property of the 3D SUSY NLSM 
holds true in all orders (in the $1/n$ expansion). We would then 
have 3D finite field theories in addition to 2D and 4D finite field 
theories, $N=4$ SUSY NLSM and Yang-Mills theories, respectively.

\bigskip
\section{Noncommutative field theories}

\subsection{Noncommutative field theories}
Noncommutative ($d-1$)-dimensional space ($d$-dimensional 
space-time) is defined by introducing the commutation relation  
among the coordinates $\hat{x}^{\mu}$, 
\beq
[\hat{x}^{\mu}, \hat{x}^{\nu}] = i \theta^{\mu \nu}, 
\eeq
where
\beq
\mu = 1, 2, \cdots, d-1 \ (\mu = 0, 1, \cdots, d-1).
\eeq
Field theories can be constructed on this space (space-time).
These noncommutative field theories have a few distinct 
properties.

\noindent
{\bf A. Perturbative}

i) Noncommutative field theories are non-local field theories
in the sense that higher derivative terms are contained in the
Lagrangian. In spite of this apparently undesirable nature,
the theories are possibly renormalizable. 4D Wess-Zumino
model is an example of consistent renormalizable
noncommutative field theories \cite{Rivel}. Renormalizability
of noncommutative extensions of 3D NLSM in $1/n$
expansions has also been investigated \cite{Gomes et al}.

ii) There appear new poles in amplitudes at zero momentum 
originating from UV divergences of loop integrals. This 
phenomenon is called UV/IR-mixing \cite{MinwRaamSei}. At 
higher orders in perturbation, it is difficult to separate these 
divergences from the usual UV divergences.

\medskip
\noindent
{\bf B. Non-perturbative}

\noindent
Many of noncommutative field theories have solitons and 
instantons. There are two kinds of solitons (instantons), those 
which exist in the commutative counterparts and those which do 
not exist in the commutative counterparts. We give two examples 
of the second kind. 

i) GMS solitons \quad 
Noncommutative scalar field theories in $d=2+1$ have 
solitons---GMS solitons \cite{GMS}. There cannot exist solitons in 
commutative scalar field theories in $d \geq 3$, as known as 
Derrick's theorem. This theorem is evaded in the 
noncommutative case because of higher derivative terms in the 
interaction Lagrangian.

ii) Instantons \quad Instantons can be constructed in 
noncommutative non-Abelian gauge theories in parallel with 
those in commutative gauge theories \cite{NekSch}. There are 
some differences. There exist instantons even in $U(1)$ gauge 
theory in the noncommutative case. Singularities in the moduli 
space of instantons in commutative gauge theories are resolved 
in the noncommutative gauge theories \cite{NekSch, LTY}. 
This will be discussed in sec.\ref{sec:Scatt}.

\subsection{Moyal product versus Weyl order product} 
\label{subsec:Moyel-Weyl}
Non-commutaive field theories can be formulated in two different 
ways. 

i) The space (space-time) coordinates $\hat{x}^{\mu}$ are 
treated as non-commuting operators. Fields 
$\hat{\phi}(\hat{x}^{\mu})$ are then non-commuting operators, 
and their product is an operator product. 

ii) The coordinates $x^{\mu}$ are treated as real numbers. Fields 
$\phi(x^{\mu})$ are functions but their product is defined as 
Moyal product. See (\ref{Moyal}) below.

The two schemes are equivalent and can be translated from one to 
the other. The scheme ii) is more appropriate for perturbative 
computations, and i) for studies of non-perturbative problems. 

We summarise the translation rule (Moyal-Weyl correspondence). 
We do this in the case of scalar field theory in $d=3$ as an 
illustration and for later use. We set
\begin{eqnarray}
(t, x, y) &=& (t, \vec x), \\
{}[\hat x, \hat y] &=& i \theta \quad (\theta>0). 
\label{NCspace}
\end{eqnarray}

The Weyl-Moyel correspondence is as follows. \\
\mbox{\parbox[t]{7cm}{Space coordinates are operators 
\quad $\hat{x}, \hat{y}$ \\
$$[\hat{x}, \hat{y}] = i \theta$$}}\ 
$\Longleftrightarrow$ \quad \ 
\mbox{\parbox[t]{6cm}{Real numbers \quad$x, y$ \\
$$x \star y - y \star x = i \theta$$}}

\medskip
\noindent
\mbox{\parbox[t]{7cm}{Fields are operators. \\
$$\hat{\phi}(\hat{x}, \hat{y}) = \int \frac{d^2 k}{(2\pi)^2} 
\tilde{\phi}(\vec k) e^{-i \vec k \cdot \hat{\vec x}}$$}}\ 
$\Longleftrightarrow$ \quad\ 
\mbox{\parbox[t]{6 cm}{Fields are functions. \\
$$\phi(x, y) = \int \frac {d^2 k}{(2\pi)^2}
\tilde{\phi}(\vec k) e^{-i \vec k \cdot {\vec x}}$$}}

\medskip
\noindent
\mbox{\parbox[t]{7cm}{Operator product of fields \\
$$\hat{\phi}_1 \, \hat{\phi}_2$$}}\ 
$\Longleftrightarrow$ \quad\ 
\mbox{\parbox[t]{6cm}{$\star$-product of fields \\
$$\phi_1 \star \phi_2$$}}

\medskip
\noindent
\mbox{\parbox[t]{7cm}{Trace of (composite) operators \\
$${\rm Tr}_H \hat f (\hat x, \hat y)$$}}\
$\Longleftrightarrow$ \quad\ 
\mbox{\parbox[t]{6cm}{Volume integral of functions \\
$$\int d^2x f(x, y)$$}}

\medskip
\noindent
\mbox{\parbox[t]{7cm}{Differentiation 
$$\hat{\partial}_x \hat{\phi} = i \theta^{-1} 
[\hat y, \hat{\phi}]$$}}\ 
$\Longleftrightarrow$ \quad\ 
\mbox{\parbox[t]{6cm}{Differentiation 
$$\partial_x \phi (x, y)$$}}\\ 

In the above formulae,  $\star$ is the Moyal product defined as 
\beq
(f \star g)(x, y) = {\rm exp} (\frac12 i \theta(\partial_{x_1} 
\partial_{y_2} - \partial_{y_1} \partial_{x_2})) 
f(x_1, y_1) g(x_2, y_2) 
\vert_{x_1 = x_2 = x, \,y_1 = y_2 = y}.
\label{Moyal}
\eeq
\noindent
There is a useful formula relating the operator and the function:
\beq
\phi(x,y)=\int du {\rm e}^{-i \theta ^{-1} yu} \langle x+
\frac u2 |\hat{\phi}(\hat{x},\hat{y}) |x-\frac u2 \rangle,
\eeq
where $\langle x | x' \rangle =\delta(x-x')$.

Note the analogy of the commutation relation 
$[\hat{x}, \hat{y}] = i \theta$ with the Heisenberg's 
commutation relation in quantum mechanics, 
\beq
[\hat{x}, \hat{p}] = i \hbar.
\eeq
The above formula of differentiation can be understood from 
this analogy. 

\subsection{Noncommutative extension of integrable field 
theories in $d=2$}
\label{subsec:Integrable}

Field theories on noncommutative space-{\it time} apparently 
have difficulty regarding unitarity and causality \cite{GomMehetc}. 
Noncommutative extension of 2D integrable field theories, if 
constructed, would be useful in clarifying this problem by solving 
the model explicitly. 

Noncommutative extension of the Wess-Zumino-Witten (WZW) 
model has been constructed \cite{FurIna, DabKLetc}, 
and its UV property has been studied at one-loop order. The 
$\beta$-function of the noncommutative $U(n)$ WZW model 
($n>1$) resembles that of the commutative WZW model at this 
order. The noncommutative $U(1)$ WZW model also has an 
infrared fixed point. We refer the reader  to \cite{FurIna} for the 
details of the computation.

We have also made an attempt at noncommutative extension of 
massive integrable models, e.g., sine-Gordon model. It is not 
difficult to construct infinitely many conservation laws modifying 
those of  the commutative sine-Gordon model. Proof of the 
integrability of the model will be completed by showing that they 
are involutive. This task is more difficult and we have not 
succeeded in this.\footnote
{Unpublished note of Furuta and Inami.} 
Integrability of noncommutative extension of sine-Gordon and 
principal chiral models has been studied in a different approach 
and the result has been presented at this workshop \cite{Mor}.

\bigskip
\section{Noncommutative $CP^n$ model in $d=3$ and BPS 
soliton solutions} 
\label{sec:NCCPn}

In commutaive space, 2D $CP^n$ model is known to be 
integrable.\footnote
{For the integrability of this model at quantum level, see the 
footnote in Sec.1.} 
Finite-action solutions of this model were constructed long time 
ago \cite{DinZak}. These solutions provide static solutions of 
the $CP^n$ model in $d=3$. In the $CP^1$ model, all static 
solutions saturate the BPS condition. In the $CP^n$ model with 
$n\geq2$, there exist non-BPS solutions \cite{DinZak}. In this 
section, we review the noncommutative extension of the $CP^n$ 
model and it's BPS solutions \cite{LeeLeeYan, FINY1}.\footnote
{Solutions of 2D noncommutative $CP^n$ model have recently 
been constructed in \cite{FodJacJon}.}

\subsection{Noncommutative $CP^n$ model in $d=3$ and BPS 
equation}

Two-dimensional noncommutative space is defined by the 
commutation relation given previously (eq.(\ref{NCspace})). We 
will use 
the complex coordinate 
$z=(x+iy)/\sqrt{2}$, and write 
\beq
[\hat{z}, \hat{{\bar z}}] = \theta.
\eeq
We use the operator formalism and introduce the creation and 
annihilation operators of the harmonic oscilator,
\beqa
\hat{a}&=&\sqrt{\theta} \hat{z},\\
{} [ \hat{a}, \hat{a}^\dag ] &=&1.
\eeqa
The basis of the Fock space is then given by
\beqa
\hat{a}|0 \big>&=&0\nonumber ,\\
|n\big>&=&\frac{(\hat{a}^\dag)^n}{\sqrt{n!}}|0\big>.
\eeqa
The differentiation and the integration are as explained in 
subsec.\ref{subsec:Moyel-Weyl}.
\beqa
\hat{\partial}_z \hat{\phi} &=& -\theta^{-1} 
[\hat{\bar{z}}, \hat{\phi} ], \quad 
\hat{\partial}_{\bar{z}}\hat{\phi} \ =\ \theta^{-1} 
[\hat{z}, \hat{\phi} ], \label{eq:diff} \\
\int d^2 x{\cal O} &\to& \rm{Tr} \hat{{\cal O}} = 2\pi\theta
\sum_{n\geq 0}\left< n| \hat{{\cal O}} | n\right>.
\eeqa

It is tedious to use the hat symbol each time to denote an operator. 
Hereafter we will omit the hat symbol, and write, e.g., $z$ instead 
of $\hat{z}$, unless confusion occurs. The hat symbol will be 
restored in sec.\ref{sec:NonBPS}.

To define the noncommutative $CP^n$ model, we take the 
($n+1$)-component vector field, which is an operator in the sense 
of subsec.\ref{subsec:Moyel-Weyl}, 
\beq
\Phi={}^{t}(\phi_1,\phi_2,\ldots,\phi_{n+1}).
\eeq
The noncommutative $CP^n$ model is defined by the Lagrangian 
\beq
L={\rm Tr}[D_\mu\Phi^\dag D^\mu\Phi
+\lambda(\Phi^\dag \Phi-1)],\label{L}\\
\eeq
where $D_\mu$ is a covariant derivative  defined by 
\beq
D_\mu=\partial_\mu\Phi-i\Phi A_\mu, 
\quad A_\mu\ =\ -i\Phi^\dag\partial_\mu\Phi,
\eeq
and $\lambda$ is the multiplier field imposing the constraint 
\beq
\Phi^\dag \Phi=1.
\eeq
This model has the global $SU(n+1)$ symmetry and $U(1)$ 
gauge symmetry, 
\beq
\Phi(x)\rightarrow \Phi(x)g(x), 
\eeq
where $g(x)\in U(1)$.

The equation of motion reads 
\beq
D_\mu D^\mu \Phi+\Phi(D_\mu\Phi^\dag D^\mu\Phi)=0. 
\label{eom0}
\eeq
The energy functional is given by 
\begin{equation}
E={\rm Tr}(|D_0 \Phi|^2+|D_z \Phi|^2+|D_{\bar{z}} \Phi|^2).
\eeq
We have the Bogomolnyi bound, 
\begin{equation}
E\ge{\rm Tr}(|D_0 \Phi|^2)+2\pi|Q|,
\end{equation}
where $Q$ is the topological charge and it is given by 
\begin{equation}
Q=\frac{1}{2\pi}{\rm Tr}(|D_z \Phi|^2-|D_{\bar{z}} \Phi|^2).
\label{Q}
\end{equation}

We look for static solutions satisfying the BPS equation  
\beqa
D_{\bar{z}}\Phi&=&0\quad \mbox{(self-dual solution)}, 
\label{SD1} \\
D_z\Phi&=&0\quad \mbox{(anti-self-dual solution)}. 
\label{SD2}
\eeqa
Self-dual (anti-self dual) solutions solve the equation of motion 
(\ref{eom0}). We also have non-BPS solutions which satisfy the 
equation of motion (\ref{eom0}) but do not satisfy the BPS 
equation. We postpone the discussion of the non-BPS solutions to 
sect.\ref{sec:NonBPS}.

\subsection{BPS solitons}
\label{subsec:BPSsoliton}

The solution of the BPS equation (\ref{SD1}) can be cast into the form 
(see 
for instance \cite{Zinn-Just}) 
\beq
\Phi=W(W^\dag W)^{-1/2},
\eeq
where $W$ is an $(n+1)$-component vector. We assume that 
$(W^\dag W)^{1/2}$ is invertible. 
It is useful to define the projection operator, 
\beq
P=W(W^\dag W)^{-1}W^\dag 
\eeq
with the properties 
\beqa
P^2\ =\ P, \quad P^\dag\ =\ P,\nonumber\\
PW=W.
\eeqa
The Lagrangian and the topological charge are  expressed in terms 
of $W$ as 
\beqa
L&=&{\rm Tr}\left[\frac{1}{\sqrt{W^\dag W}}
\partial_\mu W^\dag (1-P)\partial^\mu W
\frac{1}{\sqrt{W^\dag W}}\right],\label{WCPNLag}\\
Q&=&\frac{1}{2\pi}{\rm Tr}\left[\frac{1}{\sqrt{W^\dag W}}
(\partial_{\bar{z}}W^\dag (1-P)\partial_z W-
\partial_z W^\dag (1-P)\partial_{\bar{z}}W)
\frac{1}{\sqrt{W^\dag W}}\right].
\eeqa
The gauge transformation takes the form 
\beq
W \to W g (z,\bar{z}),
\eeq
where  $g (z,\bar{z})$ is an arbitrary function of $z$ and 
$\bar z$ which is assumed to be invertible. 

The (self-dual) BPS equation (\ref{SD1}) becomes
\begin{equation}
D_{\bar{z}}\Phi=(1-P)(\partial_{\bar{z}}W)
(W^\dag W)^{-1/2}=0.\label{SD3}
\end{equation}
This equation is equivalent to 
\beq
\partial_{\bar{z}}W=WV, 
\label{SD3}
\eeq
where $V$ is an arbitrary scalar. The general solution of 
(\ref{SD3}) is written as 
\beq
W=W_0(z)\Delta(z,\bar{z}), 
\eeq
where $\Delta(z,\bar{z})$ is a scalar. 
$W_0(z)$ is an $(n+1)$-vector  with all it's components being 
holomorphic polynomials. The degree $k$ of the polynomials 
gives the topological charge, $Q=k$. 

In the commutative case, we may set $W={}^{t}(w, 1)$ 
by a gauge transformation, $w$ being an $n$-vector whose 
components are rational functions. We will be mainly concernd 
with one- and two-soliton solutions of the $CP^1$ model. In the 
commutative $CP^1$ model, they are given by \cite{Ward, Lee} 
\beqa
w&=&\lambda+\frac{\mu}{z-\nu} 
\hspace{16.8mm}\mbox{(one-soliton solution)},\label{c1}\\
w&=&\alpha+\frac{2\beta z+\gamma}{z^2+\delta z+\epsilon}
\quad \hspace{3mm}\mbox{(two-soliton\ solution)},\label{c2}
\eeqa
where $\alpha,\beta,\cdots \in {\boldmath C}$ are parameters 
of the solutions, called {\it moduli parameters}. We may set the 
moduli parameters $\lambda$ and $\alpha$ to zero, using the 
global $SU(2)$ symmetry. 

In the noncommutative case, $W={}^{t}(z,\mu)$ and 
$W'=W z^{-1}={}^{t}(1, \mu z^{-1})$ are not gauge equivalent. 
This is because $z^{-1}$ is not invertible. Here $z^{-1}$ is 
defined by 
\beq
z^{-1}\equiv(\bar{z}z)^{-1}\bar{z}
\equiv\bar{z}(\bar{z}z+\theta)^{-1}. 
\label{inverse}
\eeq
and hence, 
\beq
z z^{-1}=1,\quad z^{-1}z=1-|0\rangle \langle 0|.
\eeq
$W$ satisfies the BPS equation but $W'$ does not. One- and 
two-soliton solutions in noncommutaive space are given by
\beqa
W&=&\left(\begin{array}{c}z-\nu\\ \mu \end{array}\right)
\hspace{13.5mm}\mbox{(one-soliton)},
\label{W1}\\\
W&=&\left(\begin{array}{c}z^2+\delta z+\epsilon\\
2\beta z+\gamma \end{array}\right)\quad
\mbox{(two-soliton)}.\label{4ts}
\eeqa

\bigskip
\section{Scattering of solitons in noncommutative $CP^1$ 
model} 
\label{sec:Scatt}

In the previous section, we have constructed static solutions of 
the 3D $CP^1$ model. We can go beyond static solutions and 
introduce time dependence to soliton solutions. It will allow 
solitons to move and scatter. The same type of problem 
already appeared in the  case of magnetic monopoles in 4D 
Yang-Mills theories. Motion of mono-poles at low energies can 
be dealt with by Manton's prescription \cite{Manton}, introducing 
time dependence to moduli parameters, and thus reducing the 
problem to geodesics in the moduli space of mono-pole solutions. 

An interesting issue regarding moduli space, especially from 
mathematical physics point of view, is the singularities of the 
instanton/soliton solutions. Short-distance singularities (in the 
moduli space) appear in commutative 4D Yang-Mills theories. 
They are shown to be resolved in the noncommutative models 
\cite{NekSch, LTY}. 

We have investigated the moduli space of the BPS solitons in 
the noncommutative $CP^1$ model from these poins of view 
\cite{FINY1}. In commutative space, the moduli space was 
investigated for one- and two-soliton solutions in \cite{Ward}. 
It was noted that there is a singularity corresponding 
to the small scale limit of the moduli parameters. 

Scattering of solitons in $d=3$ was studied previously in other 
types of 3D noncommutative scalar field theories 
\cite{LRvUetc, GHS, LecPop}.

\subsection{Low-energy dynamics and moduli space}
\label{subsec:moduli}

At low energies (near the Bogomolnyi bound), it is a good 
approximation that only the moduli parameters depend on time 
$t$. Let us denote the moduli parameters generically by 
$\alpha(t), \beta(t), \cdots$. The time evolution of the moduli 
parameters is determined by the action 
\beq
S=\int dt L[\alpha(t),\beta(t),\cdots],
\eeq 
It amounts to dealing with the kinetic energy term $T$.
\beqa
T&=&{\rm Tr}\left[\frac{1}{\sqrt{W^\dag W}}
\partial_t W^\dag (1-P)\partial^t W
\frac{1}{\sqrt{W^\dag W}}\right]\nonumber\\
&=&\frac{1}{2}{\rm{\bf Tr}}(\partial_t P)^2
,\label{Kinetic}
\eeqa
where {\bf Tr} consists of the trace over the Fock space and that 
over the 2$\times$2 matrix indices. Other terms in $L$ give the 
topological charge and thus become a constant term.

To give the metric on the moduli space, we rewrite the kinetic 
energy $T$ as follows,
\beq
T=\frac{1}{2}\left(\frac{ds}{dt}\right)^2
=\frac{1}{2}g_{ab}\frac{d\zeta^a}{dt}\frac{d\zeta^b}{dt}.
\label{Trewitten}
\eeq
Here $ds$ is a line element of the moduli space ${\cal M}$. 
$g_{ab}$ is the metric on ${\cal M}$. The moduli parameters 
$\zeta^a$ are the coordinates on ${\cal M}$. 
Eq. (\ref{Trewitten}) means that dynamics of solitons is given 
by the geodesics in ${\cal M}$. 

In the commutative $CP^n$ model, the moduli space is known to 
be a K\"ahler manifold \cite{Ruback}. We have shown that this is 
also the case in noncommutive space \cite{FINY1}. To see this, 
we rewrite the projection operator $P$ as \cite{GHS}
\beqa
P&=&\sum_{n,m}|\psi_n\rangle h^{nm} \langle\psi_m|,
\nonumber\\
& & |\psi_n\rangle=W|n\rangle,\nonumber\\
h_{nm}&=&\langle\psi_n|\psi_m\rangle,\quad h^{nm}\ =\ 
(h_{nm})^{-1}.
\eeqa
$|\psi_n\rangle$ is a holomorphic function of the moduli 
parameters. One can show that when $|\psi_n\rangle$ is a 
holomorphic function of the moduli parameters, the moduli space 
${\cal M}$ is a K\"ahler manifold with the K\"ahler potential 
\cite{GHS}
\beq
K={\rm Tr}\ln (h_{nm})={\rm Tr}\ln (W^\dag W).
\eeq
Hence we write
\begin{equation}
T=\frac{1}{2}g_{\bar{a}b}\frac{d\zeta^{\bar{a}}}{dt}
\frac{d\zeta^b}{dt},
\quad g_{\bar{a}b}=\frac{\partial}{\partial\zeta^{\bar{a}}}
\frac{\partial}{\partial\zeta^b}K.
\end{equation}

\subsection{One-soliton metric}

Recall the one-soliton solution,
\beq
W=\left(\begin{array}{c}z-\nu\\ \mu \end{array}\right)\label
{1-soliton}.
\eeq
$\nu$ represents the position of the soliton.
$|\mu|$ gives the size of the soliton. Subsitute (\ref{1-soliton}) to 
the kinetic energy (\ref{Kinetic}). Then,
\beqa
T&=&\mbox{Tr}
\Bigg[
\frac{1}{\sqrt{(\bar{z}-\bar{\nu})(z-\nu)+|\mu|^2}}\partial_t 
\left(\begin{array}{cc} \bar{z}-\bar{\nu} & \bar{\mu} 
\end{array}
\right)
\nonumber\\
& &\times\left\{1-
\left(\begin{array}{c} z-\nu \\ \mu \end{array}\right)
\frac{1}{(\bar{z}-\bar{\nu})(z-\nu)+|\mu|^2}
\left(\begin{array}{cc} \bar{z}-\bar{\nu} & \bar{\mu} 
\end{array}
\right)
\right\}\nonumber\\
& &\times\partial_t
\left(\begin{array}{c} z-\nu \\ \mu \end{array}\right)
\frac{1}{\sqrt{(\bar{z}-\bar{\nu})(z-\nu)+|\mu|^2}}
\Bigg].
\eeqa

We look at the $\dot{\bar{\mu}}\dot{\mu}$ term in $T$. 
\beq
2\pi\theta\dot{\bar{\mu}}\dot{\mu} \sum_{n\ge 0}
\frac{1}{\theta n+|\mu|^2}\left[\frac{\theta n}
{\theta n+|\mu|^2}+\frac{|\nu|^2}{\theta(n+1)+|\mu|^2}
\right].\label{div}
\eeq
The sum of the first terms in (\ref{div}) diverges. The low-energy 
aproximation would fail unless $\dot{\mu}=0$. In the 
commutaitve case, this is a typical property of the $CP^n$ model. 
The moduli space is restricted to a lower dimensional submanifold.
Non-commutativity does not change this situation. We set $\mu$ 
to a constant. 

Then we obtain,
\begin{equation}
T=2\pi\frac{d\bar{\nu}}{dt}\frac{d\nu}{dt},
 \quad \mbox{or} \quad ds^2=4\pi d\bar{\nu}d\nu.
\label{metric1}
\end{equation}  
This equation means that the geodesic is a straight line in the 
$\nu$ plane. Soliton moves straight without changing its size 
$|\mu|$.

\subsection{Two-soliton metric}

Recall the BPS two-soliton solution (\ref{4ts}),
\begin{equation}
W=\left(\begin{array}{c}z^2+\epsilon \\ 2\beta z+\gamma
\end{array}\right).\label{5ts}
\end{equation}
Here we concider the center-of-mass frame, setting 
$\delta$ to $0$ in (\ref{4ts}). 
Computing the kinetic energy in the low-energy limit, we find that
the contribution of the $\dot{\bar{\beta}}\dot{\beta}$
term diverges in the same way as $\dot{\bar{\mu}}\dot{\mu}$ 
term in the one-soliton case. In the low-energy approximation, we 
should set $\beta$ to a constant. Furthermore, for the sake of 
simplicity, we consider the case of $\beta=0$. In this case, 
$\pm i\epsilon^{1/2}$ represent the locations of the solitons. 
$|\gamma/\epsilon^{1/2}|$ is the size of the solitons (same size 
for both solitons). 

In the commutative model, the moduli space metric has been 
calculated by Ward \cite{Ward}. It is written as 
\beq
ds^2=\xi R^{-1}dR^{2}+\mu dRd\psi +\nu R d\psi^2
+R(\tau d\phi^2+\sigma d\theta d\phi +\omega d\theta^2),
\label{Wmet}
\eeq 
where $\gamma=Re^{i\phi}\sin{\psi}$, $\epsilon=Re^{i\theta}
\cos{\psi}$. 
$\xi, \mu, \nu, \tau, \sigma$ and $\omega$ are 
functions of $\psi$ only. 
The explicit forms of these functions are
\beqa
\xi=\frac{1}{2}E\ ,\quad \mu=\tan{\psi}(K-E)\ ,\quad 
\nu=K-\frac{1}{2}E\ ,
\nonumber\\
\tau=\nu\sin^2 \psi\ ,\quad\sigma=-\mu\sin\psi\cos\psi\ ,
\omega=\xi\cos^2\psi\ ,
\eeqa
where $K=K(\cos\psi), E=E(\cos\psi)$ are the complete elliptic 
functions of the first and second kind, respectively. 
One can show that at $(\epsilon=0, \gamma\neq 0)$ or 
$(\epsilon\neq 0, \gamma=0)$, 
there are no singularities by suitable coordinate redefinitions. 
Since the metric (\ref{Wmet}) is homogeneous in $R$, there is a 
singularity at $(\epsilon, \gamma)=(0,0)$, otherwise the 
whole space must be flat.  We will now see the disappearance of 
this singularity in the noncommutative model.

We have obtained the moduli space metric in the noncommutative 
model, 
\beqa
g_{\bar{\gamma}\gamma}&=&\mbox{Tr}\left[\frac{1}
{\bar{\gamma}\gamma+(\bar{z}^2+\bar{\epsilon})
(z^2+\epsilon)}
\left(1-\frac{\bar{\gamma}\gamma}{\bar{\gamma}\gamma
+(\bar{z}^2+\bar{\epsilon})(z^2+\epsilon)}\right)\right],
\label{gg}\\
g_{\bar{\epsilon}\gamma}&=&-\mbox{Tr}\left[\bar{\gamma}
(z^2+\epsilon)\frac{1}{[\bar{\gamma}\gamma
+(\bar{z}^2+\bar{\epsilon})(z^2+\epsilon)]^2}\right],
\label{ge}\\
g_{\bar{\gamma}\epsilon}&=&-\mbox{Tr}\left[\gamma\frac{1}
{[\bar{\gamma}\gamma+(\bar{z}^2+\bar{\epsilon})
(z^2+\epsilon)]^2}(\bar{z}^2+\bar{\epsilon})\right],
\label{eg}\\
g_{\bar{\epsilon}\epsilon}&=&\mbox{Tr}\left[\frac{1}
{\bar{\gamma}\gamma+(\bar{z}^2+\bar{\epsilon})
(z^2+\epsilon)}\frac{\bar{\gamma}\gamma}
{\bar{\gamma}\gamma+(\bar{z}^2+\bar{\epsilon})
(z^2+\epsilon)+4\theta\bar{z}z+2\theta^2}\right].
\label{ee}
\eeqa
It is difficult to go further to compute the Tr for arbitrary values of 
$\theta$. We can compute the Tr and obtain the 
moduli space metric explicitly in the two limiting cases: 
large values of $\theta$ and small values of $\theta$. 

\noindent
(1) $|\gamma|, |\epsilon| \ll \theta$ case.

We obtain
\begin{equation}
ds^2=\frac{2\pi}{\theta}\left(d\bar{\gamma}d\gamma
+\frac{2}{3}d\bar{\epsilon}d\epsilon\right)
+O(\theta^{-2})\label{largeth}.
\end{equation}
Results are summarized as follows.
\begin{description}
\item(i)
The moduli metric turns out to be flat.

\item(ii)
Suppose that initialy, $\epsilon >0$, $\dot{\epsilon} <0$. The 
soliton locations $\pm i\epsilon^{1/2}$ move from imaginary 
axis to real axis, as shown schematically in fig.\ref{scattering}. 
Note that this does not mean right-angle scattering since the 
expression (\ref{largeth}) holds only when the soliton locations 
are close, i.e., $|\epsilon| \ll \theta$.

\FIGURE{
\parbox{160mm}{
\begin{center}
\epsfig{file=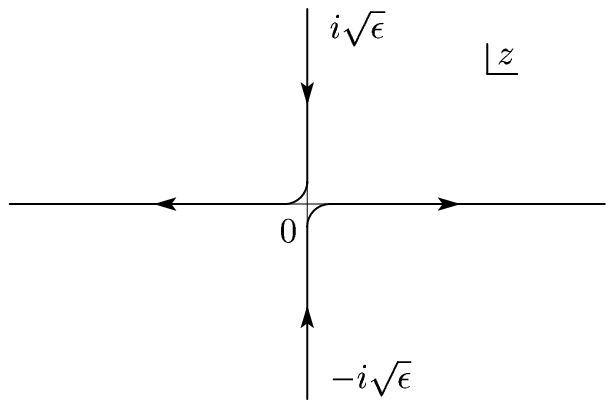}
\end{center}}
\caption
{Scattering of solitons: Two solitons going in from the 
imaginary axis $\pm i\sqrt{\epsilon} \,(\epsilon >0) $ come out to 
the real axis $\pm \sqrt{-\epsilon} \,(\epsilon <0)$.}}
\label{scattering}

\item(iii)
The singularity which exists in the commutative model at
$(\epsilon, \gamma)=(0,0)$ is resolved in the noncommutaive 
model.
\end{description} 

\noindent
(2) $|\gamma|, |\epsilon| \gg \theta$ case.

In this case, the metric can be calculated using the 
$\star$-product formalism rather than the operator formalism. 
We find
\beq
(g_{ab})_{\rm NC}=(g_{ab})_{\rm C}+O(\theta),\label{smallth}
\eeq
where $a,b$ represent the moduli parameters. NC and C denote 
noncommutative and commutative cases respectively. From 
(\ref{smallth}) one sees that the metric has the smooth 
commutative limit $\theta\to 0$ with moduli parameters fixed.

\bigskip
\section{Non-BPS solitons in noncommutative $CP^1$ model} 
\label{sec:NonBPS}

We have so far considered BPS solutions. In general there can be  
non-BPS solutions, i.e., solutions of the equation of motion which 
do not satisfy the BPS equations. It is known that there exists no 
{\it static} non-BPS solutions in the commutative 3D $CP^1$ moel 
\cite{DinZak}. We will show in this section that there exist 
static non-BPS solutions in the noncommutative 3D $CP^1$ 
model. We present solutions corresponding to soliton anti-soliton 
configurations and solutions of other types derived in \cite{FINY2}.

\subsection{Preliminary}

We rewrite some of the formulae given previously. In terms of the 
projection operator $P$ introduced in sec.\ref{subsec:BPSsoliton}, 
the Lagrangian is written as
\beq
L=\frac 12 {\rm{\bf Tr}}(\partial_t P\partial_t
P - 2\hat{\partial}_{\bar{z}}P \hat{\partial}_z P),
\label{PL}
\eeq
where {\bf Tr} is defined in sec.\ref{subsec:moduli}. The equation 
of motion is
\beq
[\partial_t^2 P - 2\hat{\partial}_{\bar{z}} \hat{\partial}_z P, 
P]=0.
\label{Peqmotion}
\eeq

We deal with static configurations. Using the differentiation
 formula (\ref{eq:diff}), eq. (\ref{Peqmotion}) is written 
as
\beq
[[\hat{z},[\hat{\bar{z}},P]],P]=0,
\label{staticPeqmotion}
\eeq
The BPS equations (\ref{SD1}), (\ref{SD2}) are now written as
\beqa
&&(1-P)\hat{z}P=0\quad\mbox{(self-dual solution)},
\label{Psolitoneq}\\
&&(1-P)\hat{\bar{z}}P=0\quad\mbox{(anti-self-dual solution)}.
\label{Pantisolitoneq}
\eeqa
Note that solutions of the BPS equations automatically satisfy the
equation of motion (\ref{staticPeqmotion}).

\subsection{Soliton-antisoliton solution and its stability}

As a candidate for non-BPS solutions, we consider the 
$2\times 2$ projection operator of the form
\beq
P=\left(
\begin{array}{cc}
P_1&0\\
0&P_2
\end{array}
\right).
\label{Pdiagonal}
\eeq
We take $P_1$ and $P_2$ which are self-dual and anti-self-dual
solutions, respectively. 
\beqa
&&(1-P_1)\hat{z}P_1=0,
\label{P1solitoneq}\\
&&(1-P_2)\hat{\bar{z}}P_2=0.
\label{P2antisolitoneq}
\eeqa
$P$ satisfies the equation of motion (\ref{staticPeqmotion}),
since $P_1$ and $P_2$ do.
$P$ does not satisfy the self-dual equation (\ref{Psolitoneq})
(anti-self-dual equation (\ref{Pantisolitoneq})),
since $P_2$ does not ($P_1$ does not).
Hence, $P$ is a desired non-BPS solution.

Non-trivial solutions of (\ref{P1solitoneq}) are known as GMS 
solitons and are given in \cite{GHS, LecPop, HLRvU}. Since $1-P_2$ 
satisfies the self-dual equation, solutions of (\ref{P2antisolitoneq}) 
are also given by GMS solitons. Therefore, $P$ takes the following 
form
\beq
P=\left(
\begin{array}{cc}
\sum_{i,j=1}^r|z^i\rangle h_{ij}^{-1}\langle z^j|&0\\
0&1-\sum_{k,l=1}^s|\tilde{z}^k\rangle\tilde{h}_{kl}^{-1}
\langle\tilde{z}^l|
\end{array}
\right),
\label{Psolitonantisoliton}
\eeq
where
\beqa
&&|z^i\rangle=e^{\theta^{-1}(z^i\hat{\bar{z}}-
\bar{z}^i\hat{z})}|0\rangle,\label{z^i}\\
&&h^{ij}=\langle z^i|z^j\rangle,
~~~h_{ij}^{-1}h^{jk}={\delta_i}^k.
\label{h}
\eeqa
A natural interpretation of the solution (\ref{Psolitonantisoliton}) 
is that it represents a soliton-antisoliton configuration with $Q=r-s$ 
and $E=2\pi(r+s)$, where $r$ and $s$ are the numbers of solitons 
and anti-solitons, respectively. Then, $z^i~(i=1,\dots,r)$ and 
$\tilde{z}^k~(k=1,\dots,s)$ are the positions of solitons and 
anti-solitons, respectively.

We analyze the stability of the solution of one soliton-antisoliton 
pair 
\beq
P=\left(
\begin{array}{cc}
|z\rangle\langle z|&0\\
0&1-|0\rangle\langle0|
\end{array}
\right).
\label{Ppair}
\eeq
We can find a path which connects this solution to the vacuum 
solution
\beq
P_0=\left(
\begin{array}{cc}
0&0\\
0&1
\end{array}
\right).
\label{Pvacuum}
\eeq
One way of parametrizing the path is given by 
\beq
P_\phi=\left(
\begin{array}{cc}
\sin^2\phi|z\rangle\langle z|&\sin\phi\cos\phi|z\rangle
\langle0|\\
\sin\phi\cos\phi|0\rangle\langle z|&1-\sin^2\phi|0\rangle
\langle0|
\end{array}
\right),
~~~\phi\in\left[0,\frac{\pi}{2}\right].
\label{Pphi}
\eeq
The energy of this configuration is
\beq
E=4\pi\sin^2\phi\left(1+\frac{\bar{z}z}{\theta}
\cos^2\phi\right).
\label{Ephi}
\eeq

It is easy to see  that the stability of the solution (\ref{Ppair}) 
depends on the separation $\vert z \vert$ between the soliton 
and the anti-soliton. When $\bar{z}z<\theta$ the energy 
(\ref{Ephi}) has a local maximum at $\phi=\frac{\pi}{2}$ and 
decreases monotonically to zero at $\phi=0$. In this case the 
solution (\ref{Ppair}) is unstable and the soliton-antisoliton pair 
annihilates. When $\bar{z}z>\theta$ the energy (\ref{Ephi}) 
has a local minimum at $\phi=\frac{\pi}{2}$ and therefore the 
solution (\ref{Ppair}) is metastable in this parameter space. We 
do not know whether the solution is unstable under fluctuations 
in other directions.

\subsection{Time-dependent solution}

Time-dependent solutions can be obtained by a boost accompanied 
by rescaling of the noncommutative parameter because the 
Lorentz symmetry is explicitly broken by the noncommutativity 
\cite{BL}. For the solution of the diagonal form (\ref{Pdiagonal}),
$P_1$ and $P_2$ can be boosted with arbitrary velocities $v_1$ 
and $v_2$. Boosted solutions take the same form as 
(\ref{Psolitonantisoliton}) but the coordinates $\hat{z}$ and 
$\hat{\bar{z}}$ are replaced by the boosted coordinates 
$\hat{z}_a$ and $\hat{\bar{z}}_a$ ($a=1,2$) which obey the 
commutation relation
\beq
[\hat{z}_a,\hat{\bar{z}}_a]=\theta_a,
~~~\theta_a=\frac{\theta}{\sqrt{1-v_a^2}},
~~~a=1,2.
\label{zaLorentzcommutation}
\eeq

Note that time does not commute with spatial coordinates due to 
the boost but static solutions are the same as those with time 
being commutative. 
The solutions constructed in this way do not represent 
time-dependent multi-(anti-)soliton configurations. For, all (anti-)
soliton peaks of our solutions move with a common velocity, 
whereas in time-dependent solutions (anti-)soliton peaks should 
exhibit relative motion.

\subsection{Non-BPS solutions of other types}

We can construct other non-BPS solutions of the form 
(\ref{Pdiagonal}). For example, we construct the non-BPS
solution
\beq
P=\left(
\begin{array}{cc}
|n\rangle\langle n|&0\\
0&1
\end{array}
\right),
~~~n>0.
\label{Pother}
\eeq
This configuration has the topological charge $Q=1$ and the 
energy $E=2\pi(2n+1)$. We do not know whether the solution 
(\ref{Pother}) can be interpreted as a soliton-antisoliton 
configuration.

As mentioned in section 4.2, $W'= {}^t(\mu\hat{z}^{-1}, 1)$ is 
not a BPS solution. Moreover, $W'$ is not a solution of the equation 
of motion. We can construct a solution by adding the correction to 
the projection operator
$P'=W'(W'^\dag W')^{-1}W'^\dag$.
Consider the projection operator
\beq
P=P'
+\frac{1}{\mu^2+\theta}\left(
\begin{array}{cc}
\theta|1\rangle\langle1|
&-\mu\sqrt{\theta}|1\rangle\langle0|\\
-\mu\sqrt{\theta}|0\rangle\langle1|
&\mu^2|0\rangle\langle0|
\end{array}
\right).
\label{Pnonbps}
\eeq
We have shown that $P$ is a non-BPS solution which has the 
topological charge $Q=1$ and the energy $E=6\pi$. The 
parameter $\mu$ is related to the size of the solution. In the 
limit of $\mu\to 0$, (\ref{Pnonbps}) reduces to
\beq
P=\left(
\begin{array}{cc}
|1\rangle\langle 1|&0\\
0&1
\end{array}
\right).
\label{Psmall}
\eeq
This corresponds to (\ref{Pother}) with $n=1$. On the other hand, 
in the limit of $\mu\to\infty$, (\ref{Pnonbps}) reduces to
\beq
P=\left(
\begin{array}{cc}
1-|0\rangle\langle 0|&0\\
0&|0\rangle\langle 0|
\end{array}
\right).
\label{Plarge}
\eeq
This corresponds to the solution representing a soliton-antisoliton 
pair sitting at the origin. We can interpret the non-BPS solution 
(\ref{Pnonbps}) as the configuration which contains a soliton of 
the size $\mu$ and a small soliton-antisoliton pair. In the large 
$\mu$ limit the soliton spreads over the space and disappears, 
and hence only the soliton-antisoliton pair exists.

\acknowledgments
This article is based on a few collaborational papers on NLSM in 
$d=3$ and one in $d=2$. We owe much to Yorinori Saito for the 
work \cite{ISY1, ISY2} and to Hiroaki Nakajima for the work 
\cite{FINY1, FINY2}. We wish to thank Koji Hashimoto for a 
valuable comment on non-BPS solutions.
This work is supported partially by 
National Science Council and National Center for Theoretical 
Science, Taiwan, ROC, 
Korea Research Foundation 2002-070-C00025, 
Research grant of Japanese Ministry of Education and Science, 
Kiban C (2) 14540265, and 
Chuo University grant for special research.

\end{document}